\documentclass{pasa}%

\usepackage{graphicx}

\title[SETI with the MWA]{A SETI Survey of the Vela Region using the Murchison Widefield Array: Orders of Magnitude Expansion in Search Space}

\author[Tremblay \& Tingay]{Tremblay, C.D.$^1$ \& Tingay, S.J.$^2$
\affil{$^1$CSIRO Astronomy and Space Science, PO Box 1130, Bentley WA 6102, Australia}%
\affil{$^2$International Centre for Radio Astronomy Research, Curtin University, Bentley, WA 6102, Australia}
}%

\jid{PASA}
\doi{10.1017/pas.\the\year.xxx}
\jyear{\the\year}

\usepackage{aas_macros}
\usepackage{hyperref} 
\hypersetup{colorlinks,citecolor=blue,linkcolor=blue,urlcolor=blue}

\hypersetup{draft}

\begin{document}

\begin{frontmatter}
\maketitle

\begin{abstract}
Following the results of our previous low frequency searches for extraterrestrial intelligence (SETI) using the Murchison Widefield Array (MWA), directed toward the Galactic Centre and the Orion Molecular Cloud (Galactic Anticentre), we report a new large-scale survey toward the Vela region with the lowest upper limits thus far obtained with the MWA. Using the MWA in the frequency range 98-128\,MHz over a 17\,hour period, a 400\,deg$^{2}$ field centered on the Vela Supernova Remnant was observed with a frequency resolution of 10\,kHz. Within this field there are six known exoplanets. At the positions of these exoplanets, we searched for narrow band signals consistent with radio transmissions from intelligent civilizations. No unknown signals were found with a 5$\sigma$ detection threshold. In total, across this work plus our two previous surveys, we have now examined 75 known exoplanets at low frequencies.  In addition to the known exoplanets, we have included in our analysis the calculation of the Effective Isotropic Radiated Power (EIRP) upper limits toward over 10\,million stellar sources in the Vela field with known distances from $Gaia$ (assuming a 10\,kHz transmission bandwidth).  Using the methods of \citet{2018AJ....156..260W} to describe an eight dimensional parameter space for SETI searches, our survey achieves the largest search fraction yet, two orders of magnitude higher than the previous highest (our MWA Galctic Anticentre survey), reaching a search fraction of $\sim2\times10^{-16}$. We also compare our results to previous SETI programs in the context of the EIRP$_{\rm min}$ - Transmitter Rate plane.  Our results clearly continue to demonstrate that SETI has a long way to go.  But, encouragingly, the MWA SETI surveys also demonstrate that large-scale SETI surveys, in particular for telescopes with a large field-of-view, can be performed commensally with observations designed primarily for astrophysical purposes.
\end{abstract}

\end{frontmatter}

\section{INTRODUCTION }
\label{sec:intro}

In this paper, we continue to report on our program to utilise the Murchison Widefield Array (MWA: \citet{Tingay_2013,Wayth_PhaseII}) in a Search for Extraterrestrial Intelligence (SETI) at low radio frequencies, over extremely wide fields of view.  

In previous work, we have examined two survey fields, encompassing 400 deg$^{2}$ toward the Galactic Centre in the frequency range 103 - 133 MHz \citep{Tingay_2016} and 625 deg$^{2}$ toward the Galactic Anticentre direction in the frequency range of 99 $-$ 122 MHz \citep{Tingay_2018}.  In these two survey fields, 45 and 22 exoplanets were known at the times of observation, respectively, and no candidate signals were detected above the observational detection limits, which were approximately $4\times10^{13}$ W and $1\times10^{13}$ W for the closest exoplanets in the fields, respectively (assuming isotropic transmitters and a 10 kHz transmission bandwidth to calculate Effective Isotropic Radiated Power: EIRP).  

A general improvement in our data processing techniques between the two sets of observations, and the fact that on average the known exoplanets toward the Galactic Anticentre are closer than those known toward the Galactic Centre, means that our upper limits on the EIRP for exoplanets toward the Galactic Anticentre are lower, in general. \cite{Tingay_2018} placed our results to that point in the context of the overall SETI endeavor and we refer the reader to that discussion and references therein for this context.

The MWA provides a unique facility to search for technosignatures at low radio frequencies, being highly sensitive, located at the radio-quiet Murchison Radio-astronomy Observatory (MRO), and having a very wide field-of-view (the surveyed areas noted above represent single MWA pointings).  The Galactic Centre field survey we previously reported was placed in the context of past SETI surveys by \cite{Gray_2017}, who show that the limits we achieved are highly competitive.  In their analysis of ``How Much SETI Has Been Done? Finding Needles in the n-dimensional Cosmic Haystack'', \citet{2018AJ....156..260W} examine an eight dimensional parameter space for radio SETI and find our two previous surveys to have the highest searched fractions for this parameter space for single surveys, factors of approximately two and ten greater than the next highest, respectively. However, the highest search fractions still sit at an order of $10^{-18}$, indicating that only a vanishingly small fraction of the SETI parameter space has been covered thus far.

While this conclusion may appear discouraging, cause for encouragement comes from the fact that SETI surveys can increasingly be performed effectively as commensal science in parallel with primary astrophysical investigations.  This has been our approach using the MWA, whereby we utilise data collected and processed in wide field searches for low frequency spectral lines (e.g. \citealt{Tremblay_2018}). The $FAST$ collaboration intends to complete commensal and dedicated SETI experiments, using a real-time data processing pipeline originally developed for the $SETI@Home$ platform to search for technosignatures from 1-1.5\,GHz during normal science operations \citep{Zhang_FAST_2020, Li_Fast}. Similar ideas exist for commensal searches being planned with MeerKAT \citep{Gajjar_2019}.

We continue this approach with the MWA here, adding a survey field centred on the Vela Supernova Remnant.  In \S \ref{sec:obs} we describe the observations and data processing, including increments in the quality of the data processing that lead to almost an order of magnitude improvement in our flux density sensitivity with commensurate improvements in our EIRP detection limits (for a fixed distance).  In \S \ref{sec:res} we describe our results, examining the six known exoplanets in the Vela field as well as the full population of stellar systems in the field (millions of systems).  In \S \ref{sec:dis} we discuss our results and conclusions.
 
\section{OBSERVATIONS AND DATA PROCESSING}
\label{sec:obs}

\begin{figure}
\centering
\includegraphics[width=0.48\textwidth]{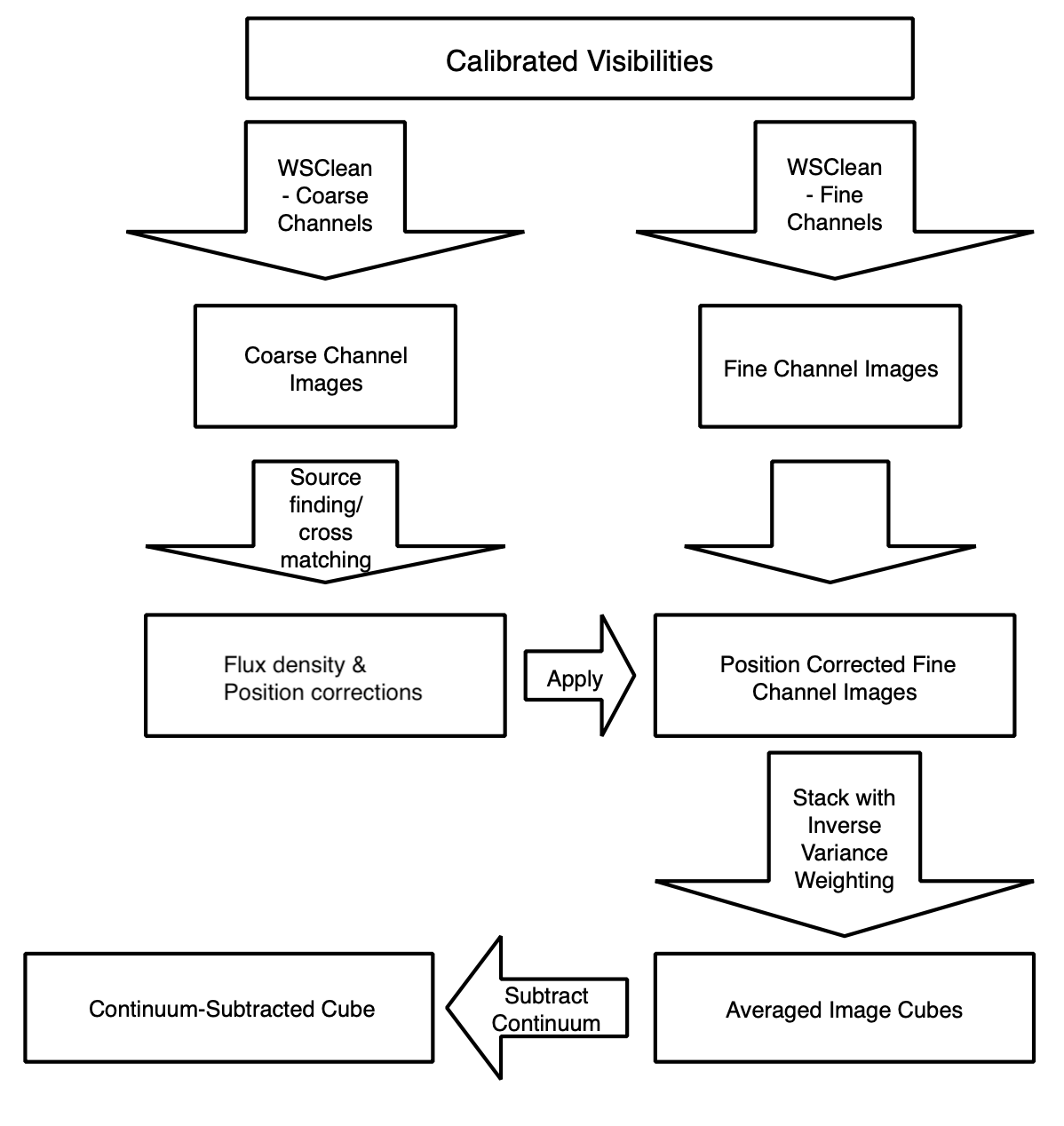}
\caption{Summary of the data processing pipeline from Figure 2 of \cite{Tremblay_2017} used to create integrated spectral cubes with the MWA.}  \label{process}
\end{figure}

\begin{figure*}
\centering
\includegraphics[width=0.90\textwidth]{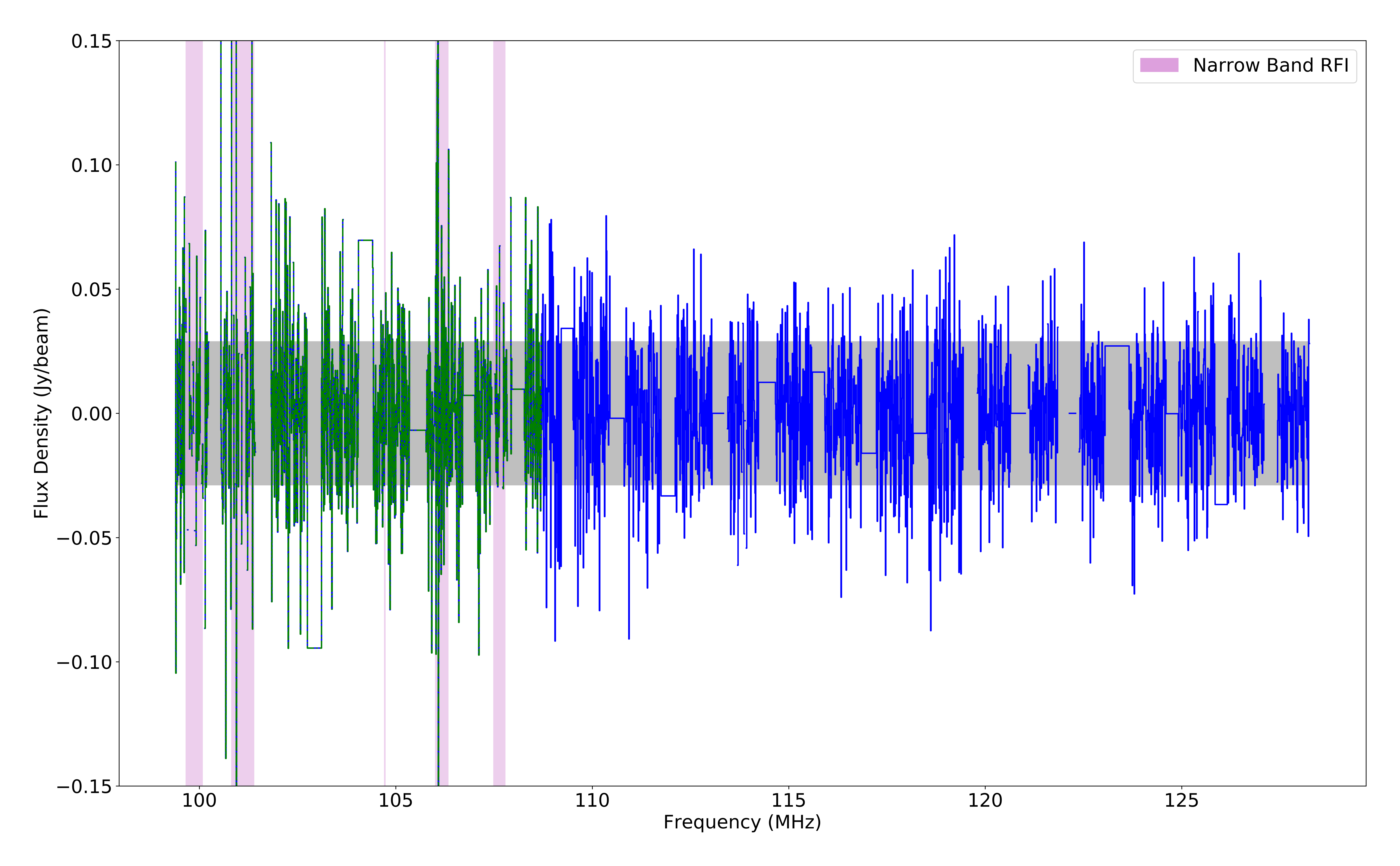}
\caption{MWA spectrum for a data cube, with a total integration time of 17 hours, used within this survey at the position of HD 75289 b. Some of the channels in the lower end of the band are affected by narrow-band RFI. The vertical shaded areas mark regions of known narrow-band RFI and the green region of the spectrum shows the top end of the FM band. Flagged channels are blanked out in the spectrum. The horizontal grey shaded region represents the $\pm1\sigma$ RMS value used in Table \ref{tab2}.}\label{spectra}
\end{figure*}

The Murchison Widefield Array (MWA; \citealt{Tingay_2013}) is a low-frequency interferometer operating between 70 and 300\,MHz at the Murchison Radio-astronomy Observatory in Western Australia.  In 2018, the telescope was upgraded to the ``Phase II'' array \citep{Wayth_PhaseII}, doubling the number of aperture array tiles from 128 to 256 and approximately doubling the maximum baseline from 3 km to 5.5 km.  

Observations of the Vela region took place between 05 January 2018 and 23 January 2018 for a total of 30\,hours, the details of which are summarised in Table \ref{obs}.  These observations were taken during the building and commissioning of the Phase II array and included 91 of the new 128 tiles.  Of the 30\,hours of observation, 17 hours were free from imaging artifacts likely caused due to the instrument being actively worked on during the day, while the observations were taken at night.

The MWA has an instantaneous bandwidth of 30.72\,MHz that is distributed between 3072 $\times$ 10\,kHz fine frequency channels.  Our data were processed following the procedure detailed in \cite{Tremblay_2017} and \cite{Tremblay_PhD} but a summary is provided here and in Figure \ref{process}.  The bandpass and phase solutions were derived each night from a two-minute observation of Hydra A (a LINER galaxy with a flux density of 243\,Jy at 160\,MHz (\citealt{Kuehr_1981})). The solutions were further refined by using self-calibration before they were applied to each 5-minute observation of the Vela region field. 

For each of the 5-minute observations, the fine frequency channels (10\,kHz) are imaged at a rate of 100 fine channels per every 1.28\,MHz coarse channel to avoid channels affected by aliasing.  This means only 78\,per\,cent of the band is imaged.  The  Phase-II  configuration  of  the MWA  used  in  these  observations  removed  the  compact  core  and  had  shortest baselines  of  1.5\,km.   In  order  to  obtain  as  much  sensitivity  to   diffuse  emission  as  possible,  all   images  were created using a Briggs weighting of 0.5. This produced a field-of-view of 400 deg$^{2}$ and a synthesised beam-width of 1$'$.  

In previous SETI surveys completed with the MWA toward the Galactic Centre and the Orion Molecular cloud, only 4 hours and 3 hours, respectively, of observations were obtained.  In this survey, 17\,hours of total integration time is used to provide our deepest low-frequency survey, producing a mean spectral RMS (root mean squared) of 0.05\,Jy\,beam$^{-1}$ across much of the field, in comparison to the previous 0.35\,Jy\,beam$^{-1}$ RMS.  

The MWA is situated in an RFI-protected environment but occasional intermittent interference occurs \citep{OffriingaRFI,Sokolowski_17}.  Each 5-minute observation was flagged using {\sc AOFlagger} \citep{OffriingaRFI} to remove strong radio frequency interference (RFI) signals from the raw visibilities based on statistical methods.  This is not expected to impact our science goals, as the chance of a real astronomical or signal from an ETI being strong enough to be flagged in a single 5-minute observation is very small.  It is estimated that this process removes less than 5--20\,per\,cent of the total visibilities, having little impact on an observation's sensitivity. For these observations, after integrating each of the snapshot images, significant narrow-band RFI was detected in the commercial FM radio bands between 98 MHz and 108\,MHz.  This left 64\% of the band available for narrow band signal searches.

These data are comensally searched for spectral line signals of an astrophysical nature, which will be reported in a separate publication (Tremblay et al. ApJ submitted).  An example of a typical spectrum with no significant signal, as seen toward HD 75289 b, is shown in Figure \ref{spectra}. 

The source finding software {\sc Aegean} \citep{Hancock_Aegean} is used to search each of the 2400 (10\,kHz) fine frequency channels independently for signals over a 5$\sigma$ limit. {\sc Aegean} works by fitting Gaussians to the pixel data and applies a correction for the background\footnote{The background is defined by the 50$^{th}$ percentile of flux distribution in a zone 30 times the size of the synthesised beam.} to calculate the flux density for potential sources. Any potential source is further evaluated based on various quality control checks, including but not limited to, ensuring the signal is greater than 5$\sigma$ in both the spectral and image plane.  Any remaining signals are cross referenced to a combination of chemical databases and new chemical modeling reported in a future publication.  Following this search, we found no signals of an unknown nature.

\begin{table}
\small
\caption{MWA Observing Parameters}

\label{obs}
\begin{tabular}{lc}
\hline
Parameter & Value\\
\hline
\hline
Central frequency& 113.28\,MHz\\
Total bandwidth & 30.72\,MHz\\
Number of imaged channels & 2400\\
Channel separation & 10 kHz \\
Synthesized beam FWHM$^*$ & 1.0$^{\prime}$\\
Primary beam FWHM$^*$ & 30\, degrees\\
Phase center of image (J2000) &  08h35m27s --45d12m19s\\
Total Integration Time & 17\,hours \\
						
\hline
$^*$Full Width at Half Maximum (FWHM)
\end{tabular}
\end{table}

\section{RESULTS}
\label{sec:res}

Utilising the processed data described in \S \ref{sec:obs}, we have access to the spectrum across our bandwidth at 10\,kHz resolution, for every pixel in our 400 deg$^{2}$ field-of-view.  Thus, as per our previous work \citep{Tingay_2018,Tingay_2016}, we can examine the spectrum at the locations of stellar systems with known exoplanets and search for narrow band signals that may constitute technosignatures.  We undertake this examination below in \S \ref{sec:known}.  Further, given that only a tiny fraction of the exoplanets in this field are known, we can also undertake a blind survey of all stellar systems in the field with known distances from \textit{Gaia} \citep{Bailer-Jones_2018} and undertake the same search.  We undertake this examination in \S\ref{sec:other}, below.

To calculate the upper limit on the total Equivalent Isotropic Radiated Power (EIRP) we use the equation:
\begin{equation}
    \mathrm{EIRP (W)} < 1.12\times10^{12} S_{\mathrm{rms}} R^{2} ,
\end{equation}
where S$_{\mathrm{rms}}$ is the RMS intensity value in Jy\,beam$^{-1}$ and $R$ is the distance to the stellar system in pc. This assumes that the transmission bandwidth is matched to the MWA fine channel bandwidth of 10\,kHz.  For transmission bandwidths less than 10\,kHz, the maximum EIRP estimates are increased by $\frac{10 kHz}{\Delta \nu_{t}}$, where $\Delta \nu_{t}$ is the transmission bandwidth.  For example, a 10\,Hz transmission bandwidth would cause our EIRP upper limits to be raised by a factor of one thousand.

\subsection{KNOWN EXOPLANETS IN THE SURVEY FIELD}
\label{sec:known}
A search of our field-of-view in the Exosolar Planets Encyclopedia Catalog\footnote{http://exoplanet.eu/catalog} (as of March 2020) returns six exoplanets hosted by five stellar systems.  These exoplanets are listed in Table \ref{tab2}, including basic parameters of the exoplanets, their stellar host, and the radio observations.  In no case were any narrow band signals detected toward these objects in our observing band at or above a level of 5$\sigma$.  As in our previous work, we assign upper limits to the EIRP based on the RMS of the measured spectrum, listed in Table \ref{tab2}.

\begin{table*}
\caption{Known exoplanets in the survey field, from the exoplanet catalog: http://exoplanet.eu/} 
\centering
\begin{tabular*}{\textwidth}{@{}c\x c\x c\x c\x c\x c\x c\x c\x c\x c@{}}
\hline \hline
Designation   & RA (J2000)   & Dec (J2000)     &  Distance  &  MSin(i) 
         & Period & Spectral & Detection  &  RMS    & EIRP$^{d}$    \\
         & hh:mm:ss        & dd:mm:ss          & (pc)                  & (M$_{J}$$^{a}$) 
         & (days) & type$^{b}$            & method$^{c}$             & (Jy/beam)                  & (10$^{13}$ W)        \\
\hline
HD 75289 b& 08:47:40.0& $-$41:44:12&28.94&0.47&3.50928&G0 V&RV&0.034&$<$3.2 \\
&&&&&$\pm$7.2946$\times10^{-5}$&&&& \\ \hline
HD 73526 b&08:37:16.0&$-$41:19:08&99.0&2.25$\pm$0.12&188.9$\pm$0.1&G6 V&RV&0.048&$<$53 \\ \hline
HD 73526 c&08:37:16.0&$-$41:19:08&99.0&2.25$\pm$0.13&379.1$\pm$0.5&G6 V&RV&0.048&$<$53 \\ \hline
HD 70642 b& 08:21:28.0& $-$39:42:19&28.8&2.0&2231$\pm$400&G5 IV-V&RV&0.039&$<$3.7 \\ \hline
DE0823-49 b& 08:23:03.0& $-$48:47:59&20.69$\pm$0.06&$-$&247.75$\pm$0.64&$-$&I&0.044&$<$2.1 \\ \hline
KELT-15 b& 	07:49:40.0&$-$52:07:14&201.0$\pm$19&0.91$\pm$0.22&3.329441&$-$&PT&0.052&$<$237 \\ 
&&&&&$\pm$1.6$\times$10$^{-5}$&&& \\
\hline \hline
\end{tabular*}\label{tab2}

\medskip
\tabnote{$^a$Mass of planet times the sine of orbit inclination, in Jupiter masses}
\tabnote{$^{b}$Spectral type of host star}
\tabnote{$^{c}$RV= Radial Velocity; I=Imaging; PT=Primary Transit}
\tabnote{$^{d}$Equivalent Isotropic Radiated Power}
\end{table*}

\subsection{OTHER STELLAR SYSTEMS IN THE SURVEY FIELD}
\label{sec:other}
\begin{figure}
\includegraphics[width=0.48\textwidth]{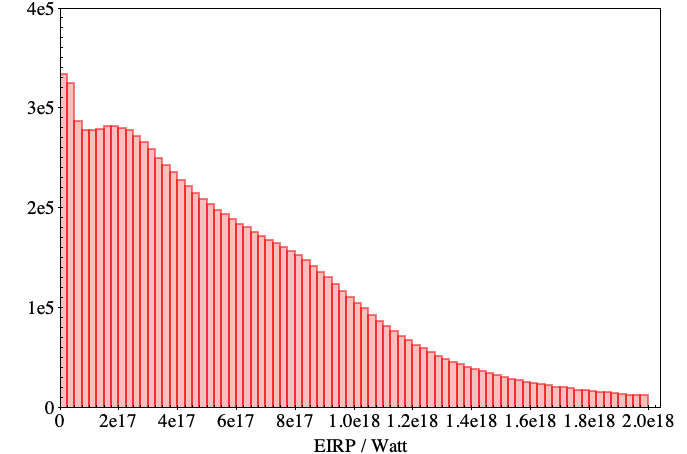}
\caption{Histogram of EIRP upper limits based on the distribution of stellar distances (set to a maximum of 6350\,pc) from the \textit{Gaia} catalog.}\label{gaia1}

\end{figure}

Given the small number of known exoplanets associated with the stellar systems in this field, it is likely that a vast number of exoplanets remain unknown.  Thus, we examine the general limits we can derive for stars in this field.  We do this by examining the \textit{Gaia} catalog, extracting the distances of stars within the field determined by their parallax measurements \citep{Bailer-Jones_2018}.  There are 10,355,066 such stars within the field-of-view for this survey. As no detections were made in the search discussed in \S \ref{sec:obs}, we utilise the RMS value as a function of position across our field and the coordinates and distances of the stars, to derive the EIRP upper limit histogram for all 10,355,066 stars with a distance smaller than 6350\,pc in Figure \ref{gaia1}.  As the distribution of distances is dominated by stars within the spiral arm of the Galaxy at 1.5--2.5\,kpc, the EIRP upper limits are also very large, orders of magnitude larger than the lowest upper limits from our previous work.  

\begin{figure}
\includegraphics[width=0.48\textwidth]{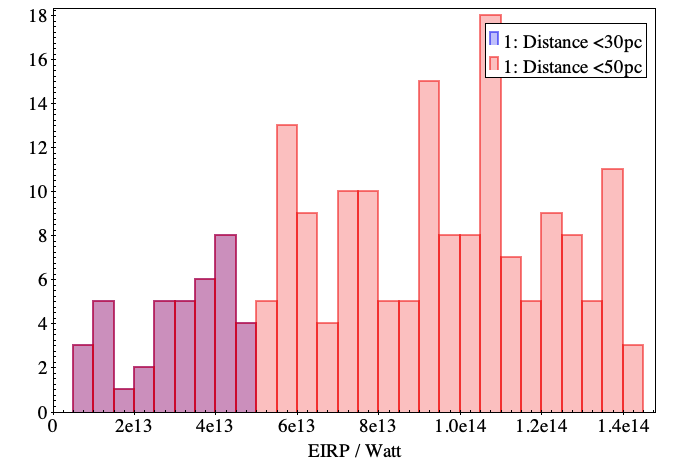}
\caption{Histogram of EIRP upper limits based on the distribution of distances from the \textit{Gaia} catalog, with a focus on sources with distances $<$50 and $<$30\,pc.}\label{gaia2}

\end{figure}

To examine the most interesting part of this distribution, at low values of EIRP upper limit, we show the EIRP upper limit histogram for those stars within 30 and 50\,pc in Fig \ref{gaia2}.  For the ten closest stars, Table \ref{tab3} lists the RMS and derived EIRP limits in more detail (excluding those systems referenced in Table \ref{tab2}).

\begin{table*}
\caption{Gaia stellar systems in the survey field, from the Gaia DR2 release} 
\centering
\begin{tabular*}{\textwidth}{@{}c\x c\x c\x c\x c\x c@{}}
\hline \hline
Designation   & RA (J2000)   & Dec (J2000)     &  Est. Distance  &  
           RMS    & EIRP$^{a}$    \\
         & hh:mm:ss        & dd:mm:ss          & (pc)   &(Jy/beam) & 10$^{13}$\,(W)      \\
\hline
5534076974490020000& 08:00:39.6& $-$41:09:59.16 & 10.58$\pm$0.26& 0.058 &$<$0.774 \\ \hline
5528866839863560000& 08:43:17.6& $-$38:52:51.30 & 11.18$\pm$0.00& 0.044  &$<$0.620 \\ \hline
5521313740446190000& 08:15:13.0& $-$42:45:40.83 & 12.60$\pm$0.27&0.039  & $<$0.733\\ \hline
5522879208777730000& 08:27:11.5& $-$44:59:13.17 & 13.76$\pm$0.01&0.033  &$<$0.698 \\ \hline
5329084752471810000& 08:44:38.3& $-$48:05:08.67 & 14.56$\pm$0.01&0.035  &$<$0.845 \\ \hline
5329751125239010000& 08:43:10.5& $-$46:59:28.18 & 15.55$\pm$0.02& 0.042 &$<$1.14 \\ \hline
5529120822749720000& 08:40:40.8& $-$38:32:38.49 & 16.04$\pm$0.02&0.036  &$<$1.06 \\ \hline
5328649002269670000& 08:53:15.0& $-$48:12:49.59 & 16.17$\pm$0.31&0.054  &$<$1.66 \\ \hline
5329580357345450000& 08:47:19.1& $-$46:52:49.80 & 16.47$\pm$0.01&0.048  &$<$1.47 \\ \hline
5514929155583860000& 08:23:02.8& $-$49:12:01.08 & 20.66$\pm$0.20&0.042  &$<$2.08 \\ \hline
\hline \hline
\end{tabular*}\label{tab3}

\medskip
\tabnote{$^{a}$Equivalent Isotropic Radiated Power}
\end{table*}

\section{DISCUSSION AND CONCLUSION}
\label{sec:dis}
The median distance for the six known exoplanet systems in the Vela field is 28.8\,pc (treating HD 73526 b and c as a single system), compared to 50\,pc for the 22 exoplanets examined by \citet{Tingay_2018} and $\approx$2\,kpc for the 45 exoplanets examined by \citet{Tingay_2016}.  Coupled with the sensitivity improvements we obtained (gained from the increased integration time of 17\,hours compared to the previous 4\,hours) described in \S \ref{sec:obs}, the median EIRP upper limit from this work is therefore an order of magnitude better than the median upper limit from \citet{Tingay_2018}.  This represents continued improvement in our techniques and general upper limits.  Our best upper limit, from Table \ref{tab3} of $6.2\times10^{12}$ W (not for a known exoplanet) approaches a 50\% improvement on our best upper limit from \citet{Tingay_2018}, still noting that an EIRP of $10^{12}$ W is high compared to the highest power transmitters on Earth at these frequencies (see \citet{Tingay_2016} for a discussion).  

\cite{Sheikh_2020} recently completed an in-depth analysis of 20 stars within the Earth transit zone between 3.95--8\,GHz with the NRAO Green Bank Telescope (GBT).  They determined an EIRP of detectable narrow-band signals that ranged from 47--17590$\times$10$^{9}$\,W for stellar distances between 7--143\,pc.  These are similar distances to the stars we present in Table \ref{tab3} for the nearby stars from $Gaia$ in the Vela field, but our sample represents a much larger population of on-average closer sources. \cite{Sheikh_2020} also convert their EIRP values to the fraction of signal capacity for the Aricebo Transmitter (L$_{A}$).  Their value of 0.033 for a star at 27\,pc is a factor of two lower than for a star in our survey at the same distance (0.068), recognising the difference in frequency between the GBT and the MWA and the fact that the GBT has better frequency resolution than the MWA.

For the first time, we obtain simultaneous upper limits on EIRP for in excess of 10 million stellar systems without known exoplanets, although the EIRP limits for the majority of distant systems are well above $10^{13}$ W.  For any future exoplanet discoveries for systems in this field, the low frequency EIRP upper limits are immediately available from our data.

\cite{Seto_2020} completed an astrometric study of F-,G- and K-type stars in $Gaia$ Data Release 2 for interstellar communications, from the view point of the sender, and concluded that surveys like $Gaia$ will be necessary to target these potential signals. \cite{Petigura_2013} suggest that approximately 20\,per\,cent of Galactic sun-like stars could have Earth-sized planets in their habitable zones and \cite{Kipping_2020} suggests that searching for technosignatures from stars with stellar types much earlier than our Sun may not be necessary, as life is unlikely to ever evolve. This means that matching SETI survey data to the $Gaia$ survey is going to be an important approach for the future of SETI.

We compare our results to Figure 5 of \cite{Price_2019}, in which the results of previous surveys are presented in a plane defined by minimum EIRP (EIRP$_{\rm min}$) at the maximum stellar distance and Transmitter Rate, (N$_{\mathrm{star}}$($\frac{\nu_{c}}{\nu_{tot}}))^{-1}$, where N$_{\mathrm{star}}$ is the total number of stars searched and $\nu_{c}$ and $\nu_{tot}$ are the central frequency of the band (113.28\,MHz) and the total bandwidth (30.72\,MHz), respectively.  For our survey using the $Gaia$ catalog, we see that our results sit below the most constraining limits set by prior work within this particular parameter space when using a distance of 1.7\,kpc and a channel bandwidth of 10\,kHz.

We also consider a different metric, utilising the method of \citet{2018AJ....156..260W} in order to calculate the ``haystack fraction'' accessible to our observations of the Vela region.  This metric takes into account the observational parameters without significant assumptions. We find the haystack fraction to be $\sim2\times10^{-16}$, which is almost two orders of magnitude higher than the highest previous fraction listed in \citet{2018AJ....156..260W}, which was for our previous observations of the Orion molecular cloud (Galactic Anticentre) field.  Our new result yields an haystack fraction almost three orders of magnitude higher than the largest non-MWA survey listed by \citet{2018AJ....156..260W}.  

Recently, \citet{Westby_2020} described the so-called Strong Astrobiological Copernican scenario, in which life must arise in a system on timescales comparable to those experienced on Earth (4.5 $-$ 5 Gyr) and posit on this basis that at least $36\pm^{175}_{32}$ civilisations capable of generating technosignatures exist in our Galaxy.  The closest system to Earth would be $17000\pm^{33600}_{10000}$ lt-yr distant.  The numbers are not large and represent a very small part of the haystack fraction parameter space, throwing into focus that SETI experiments will need to enter the statistical domain of $Gaia$-sized samples.

Overall, our MWA surveys show the rapid progress that can currently be made in SETI at radio frequencies, using wide field and sensitive facilities, but also show that SETI surveys have a long way to go.  The continued use of the MWA, and the future similar use of the SKA at much higher sensitivities, offers a mechanism to make significant cuts into the haystack fraction of \citet{2018AJ....156..260W}, while maintaining a primary focus on astrophysical investigations, making excellent commensal use of these large-scale facilities.

\begin{figure*}
\centering
\includegraphics[width=0.72\textwidth]{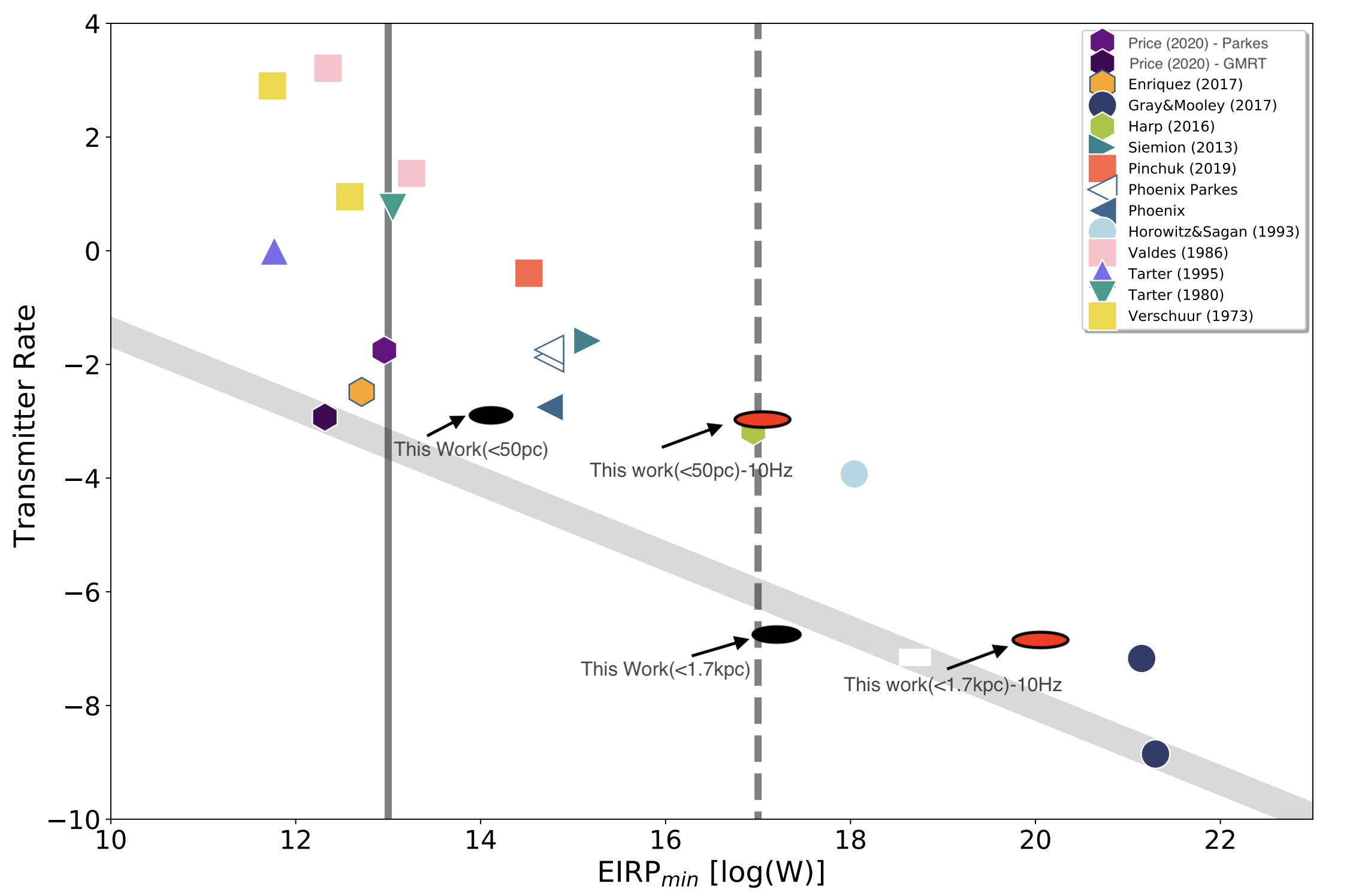}
\caption{Figure 5 from \cite{Price_2019} with our results for the $Gaia$ catalog survey (Section 3.2) shown for comparison. We report the results using a maximum RMS of 0.06\,Jy\,beam$^{-1}$ and when we limit the distances to less than 50\,pc, as per Figure \ref{gaia2} and for all stars less than 1.7\,kpc. The value of 1.7\,kpc is chosen as it is the distance to the Vela Molecular Cloud complex, a stellar rich environment toward the Carina-Sagittarius spiral arm. The black ovals are the EIRP values per Equation 1 assuming a 10\,kHz channel bandwidth and the red ovals are using the EIRP values assuming a transmission bandwidth of 10\,Hz. These results span the diagonal grey line representing a fit between the previous most constraining data points for Transmitter Rate and EIRP$_{\mathrm{min}}$ at the most distant star when using the MWA channel bandwidth.  The solid and dashed vertical lines represent the EIRP of the Arecibo planetary radar, and the total power from the Sun incident on the Earth, respectively.}
\label{Price}
\end{figure*}

\section{acknowledgements}
We would like to thank Daniel Price (Swinburne University) for his insight and comments on this paper and Natasha Hurley-Walker for help with the continuum subtraction.  We would also like to thank the anonymous reviewer for their helpful comments that significantly improved our manuscript.

\subsection {Facilities}
This scientific work makes use of the Murchison Radio-astronomy Observatory, operated by CSIRO. We acknowledge the Wajarri Yamatji people as the traditional owners of the Observatory site. Support for the operation of the MWA is provided by the Australian Government (NCRIS), under a contract to Curtin University administered by Astronomy Australia Limited.  Establishment of ASKAP, the Murchison Radio-astronomy Observatory and the Pawsey Supercomputing Centre are initiatives of the Australian Government, with support from the Government of Western Australia and the Science and Industry Endowment Fund. 
\subsection{Computer Services}
We acknowledge the Pawsey Supercomputing Centre which is supported by the Western Australian and Australian Governments. Access to Pawsey Data Storage Services is governed by a Data Storage and Management Policy (DSMP). The All-Sky Virtual Observatory (ASVO) has received funding from the Australian Commonwealth Government through the National eResearch Collaboration Tools and Resources (NeCTAR) Project, the Australian National Data Service (ANDS), and the National Collaborative Research Infrastructure Strategy. This research has made use of NASA’s Astrophysics Data System Bibliographic Services.
\subsection{Software}
The following software was used in the creation of the data cubes:
\begin{itemize}
    \item {\sc aoflagger} and {\sc cotter} -- \cite{OffriingaRFI}
    \item {\sc WSClean} -- \cite{offringa-wsclean-2014,offringa-wsclean-2017}
    \item {\sc Aegean} -- \cite{Hancock_Aegean}
    \item {\sc miriad} -- \cite{Miriad}
    \item {\sc TOPCAT} -- \cite{Topcat}
\end{itemize}


\bibliographystyle{pasa-mnras}
\bibliography{mwa-seti}

\end{document}